\documentclass[useAMS,usenatbib,usegraphicx]{mn2e}

\title[Magnetic field variability of $\gamma$ Equ]{ Secular variability 
   of the longitudinal magnetic field \\ of the Ap star $\gamma$ Equ }

\author[V.D. Bychkov, L.V. Bychkova and J. Madej]{
V.D. Bychkov$^{1}$\thanks{E-mail: vbych@sao.ru (VDB); 
lbych@sao.ru (LVB); jm@astrouw.edu.pl (JM)}, L.V. Bychkova$^{1}$ and
J. Madej$^{2}$ \footnotemark[1] \\
$^{1}$ Special Astrophysical Observatory of the Russian
       Academy of Sciences, Nizhnij Arkhyz, 369167 Russia \\
$^{2}$ Warsaw University Observatory, Al. Ujazdowskie 4,
       00-478 Warsaw, Poland }

\begin{document}

\date{Accepted .................................. }

\pagerange{\pageref{firstpage}--\pageref{lastpage}} \pubyear{2005}

\maketitle

\label{firstpage}

\begin{abstract}
We present an analysis of the secular variability of the longitudinal 
magnetic field $B_e$ in the roAp star $\gamma$ Equ (HD 201601). Measurements
of the stellar magnetic field $B_e$ were mostly compiled from the literature,
and append also our 33 new $B_e$ measurements which were obtained with the
1-m optical telescope of Special Astrophysical Observatory (Russia).
All the available data cover the time period of 58 years, and include
both phases of the maximum and minimum $B_e$. We determined that the
period of the long-term magnetic $B_e$ variations equals 
$91.1 \pm 3.6$ years, with 
$B_e({\max}) = + 577 \pm 31$ G and $B_e({\min}) = -1101 \pm 31 $ G.
\end{abstract}

\begin{keywords}
Stars: magnetic fields -- stars: chemically peculiar
              -- stars: individual: HD 201601 
\end{keywords}

\section{Introduction}

The Ap star $\gamma$ Equ (HD 201601, BS 8097) is one of the brightest
objects of this class, with the apparent luminosity $V=4.66$ mag. The exact 
spectral type of this object
is A9p (SrCrEu subclass). The magnetic field of $\gamma$ Equ has been
studied for more than 50 years, starting from October 1946 (see Babcock
1958). The longitudinal magnetic field $B_e$ of this star does not exhibit
periodic variations in time scales typical of stellar rotation,
$0.5 - 30$ days. Such a variability of the $B_e$ field was observed in most
Ap stars. The above effect is commonly interpreted as the result of
stellar rotation (oblique dipole model).

The first measurements by Babcock (1958) showed that the value of the
longitudinal magnetic field $B_e$ of $\gamma$ Equ was positive in 1946--52,
and approached nine hundred G. From that time on the value of $B_e$ slowly
decreased and even changed sign in 1970/71. One could interpret the magnetic 
behavior of $\gamma$ Equ either as secular variations, or variations
caused by extremely slow rotation. If the latter picture is correct,
then the corresponding magnetic and rotational periods are in the range
from 72 to 110 years (Bonsack \& Pilachowski 1974; Leroy et al. 1994; 
Bychkov \& Shtol' 1997; Scholz et al. 1997).

The behavior of the $B_e$ field in $\gamma$ Equ was investigated by many
authors in the second half of the twenty{\sl th} century. For this research
we compiled $B_e$ observations published by Bonsack \& Pilachowski (1974),
Scholz (1975; 1979), Borra \& Landstreet (1980), Zverko et al. (1989), 
Mathys (1991), Bychkov et al. (1991), Bychkov \& Shtol' (1997), Scholz et
al. (1997), Mathys \& Hubrig (1997), Hildebrandt et al. (2000),
Leone \& Kurtz (2003) and Hubrig et al. (2004).

We included in this paper our unpublished magnetic $B_e$ measurements which
were obtained during the past seven years. All the new magnetic observations
showed, that the slow decrease of the $B_e$ field in $\gamma$ Equ apparently
reached the minimum in 1996--2002 and has actually started to increase. 

In this paper we determined the accurate parameters of secular
variability of $\gamma$ Equ: the period $P_{mag}$, the amplitude and the
time of zero phase for $B_e$ variations, which were approximated by a sine
wave. We support the hypothesis that the long-term $B_e$ 
variation in $\gamma$ Equ is a periodic feature. Possible origin of this
variation cannot be uniquely determined, see discussion in 
Section ~\ref{sec:discussion} of this paper.

\section{ Observations and data processing }

We have performed spectropolarimetric observations of Zeeman line splitting
for $\gamma$ Equ at the Coude focus of the 1-m optical telescope (Special
Astrophysical Observatory, Russian Academy of Sciences).
Zeeman spectra were obtained with the echelle spectrograph GECS (Musaev
1996). We have put the achromatic analyser of circularly polarised light
in front of the spectrometer slit. Images of the Zeeman echelle spectra
were recorded from CCD detectors in standard FITS format.
Final reduction of the archived spectra was performed with the standard
MIDAS software (Monin 1999).

Effects of instrumental polarisation on $B_e$ measurements obtained with
this instrument were investigated by Bychkov et al. (1998, 2000).

Table~\ref{tab:saores} presents the full set of our $B_e$ measurements
of $\gamma$ Equ (total 33 $B_e$ points). 
The meaning of the first 3 columns is obvious. The fourth column
gives the number $N$ of spectral lines which were
used for the measurement of $B_e$ for a given exposure. 
Time length $\Delta t$ of the exposure (in min) is given in the last column
of Table~\ref{tab:saores}.

On average, the value of a single $B_e$ number listed in 
Table~\ref{tab:saores} was obtained after averaging of $B_e$ measurements
obtained in 500-1300 spectral lines. Standard deviation $\sigma_{B_e}$
for the resulting value of $B_e$ was computed in the standard manner as
the error of an arithmetic mean value.

Errors $\sigma_{B_e}$ determined in the above way reached rather low values
in several observations listed in Table~\ref{tab:saores}. In 2005/2006
we plan to verify the reality of such $\sigma_{B_e}$ by a special program
of $B_e$ observations. Actually we accept these errors {\sl bona fide}
and note the following properties of our $B_e$ measurements.

The referee pointed out that a few pairs of $B_e$ measurements of one night
in Table~\ref{tab:saores} differ by only a few G, which is substantially
less than the corresponding standard deviation $\sigma_{B_e}$. 
We can explain this only as a purely random effect, and do not see
any reason for it either in the acquisition of observational data or
their reduction.

Secondly, series of measurements taken within a few nights generally 
show a scatter of the order of 100 G, which is much higher than the 
standard errors $\sigma_{B_e}$ in Table~\ref{tab:saores}. The latter are 
of the order of $20-30$ G, and such a discrepancy suggests that our
standard deviations are systematically underestimated, and are in fact
of the order of 100 G. On the other hand,
such a scatter of $\approx$ 100 G is not inconsistent with the
short-term variability of light and the longitudinal magnetic field
$B_e$ in $\gamma$ Equ in time scales of minutes or above it. 

Leone \& Kurtz (2003) recently discovered periodic variations of the
longitudinal magnetic field $B_e$ in $\gamma$ Equ over the pulsation
period of this star, $P_{puls} = 12.1$ min. The estimated amplitude
$\Delta B_e = 240$ G for this period, therefore, these variations 
at least can contribute to the scatter of our $B_e$ points collected
in Table~\ref{tab:saores}. 

Study of the rapid periodic $B_e$ variations on a time scale of minutes 
was also presented in Bychkov et al. (2005b) for $\gamma$ Equ. They
did not found conclusive evidence of such variations above the noise
level at $\approx 240$ G.   

We also performed spectral analysis of the full set of 298 $B_e$ time
series from years 1946--2004. We concluded that there are no short-period
field variations with periods above ca. 1 day, but were not able to extend
our analysis for shorter periods, see Section 4 of this paper.

% \clearpage\newpage
{
\newdimen\digitwidth
\setbox0=\hbox{\rm0}
\digitwidth=\wd0
\catcode`?=\active
\def?{\kern\digitwidth}

\newcommand{\di}{\displaystyle}
\begin{table}
\caption{Measurements of $B_e$ in $\gamma$ Equ (HD 201601). }
\label{tab:saores}
\renewcommand{\arraystretch}{1.1}
\begin{tabular}{|l|r|c|r|c|}
\noalign{\vskip 2 mm}
\hline
JD \hskip1mm 2400000.+  & $B_e$ (G) & $\sigma_{B_e}$ (G)&\hskip-2mm $N$ &
\hskip2mm $\Delta t$ (min)\\
\hline
49648.323   & --1045   & 21 & 706 & 30\\[-1.0pt]
49648.345   & --1315   & 26 & 755 & 30\\[-1.0pt]
49649.229   & --1463   & 37 & 576 & 30\\[-1.0pt]
49649.257   & --1159   & 31 & 656 & 30\\[-1.0pt]
49932.424   & --1317   & 26 & 691 & 60\\[-1.0pt]
49932.469   & --1317   & 26 & 675 & 60\\[-1.0pt]
49933.460   & --1316   & 26 & 700 & 60\\[-1.0pt]
49933.507   & --1317   & 29 & 704 & 60\\[-1.0pt]
50023.158   & --1291   & 22 & 501 & 40\\[-1.0pt]
50023.189   & --1380   & 23 & 650 & 40\\[-1.0pt]
50066.128   & --1539   & 26 & 718 & 40\\[-1.0pt]
50066.157   & --1611   & 62 & 532 & 40\\[-1.0pt]
51533.1229  & --1014   & 16 & 966 & 30\\[-1.0pt]
51533.1451  & --1011   & 14 & 701 & 30\\[-1.0pt]
51535.1847  & -- 902   & 16 & 955 & 40\\[-1.0pt]
51535.2153  & -- 901   & 19 & 855 & 40\\[-1.0pt]
51536.1069  & -- 670   & 18 & 821 & 30\\[-1.0pt]
51536.1285  & -- 642   & 24 & 508 & 30\\[-1.0pt]
51888.166   & --1069   & 18 & 847 & 30\\[-1.0pt]
51888.190   & --1092   & 20 &1353 & 30\\[-1.0pt]
51889.103   & -- 890   & 20 & 847 & 30\\[-1.0pt]
51889.126   & -- 865   & 20 & 817 & 30\\[-1.0pt]
51890.142   & -- 742   & 21 & 770 & 30\\[-1.0pt]
52163.3000  & -- 845   & 19 & 833 & 30\\[-1.0pt]
52163.3201  & -- 855   & 19 & 732 & 30\\[-1.0pt]
52164.2861  & -- 956   & 16 & 947 & 30\\[-1.0pt]
52164.3076  & -- 967   & 16 & 914 & 30\\[-1.0pt]
52165.2812  & --1061   & 17 & 835 & 40\\[-1.0pt]
52165.3111  & --1029   & 16 & 991 & 40\\[-1.0pt]
52186.2229  & -- 922   & 17 &1085 & 30\\[-1.0pt]
52186.2451  & -- 942   & 17 &1055 & 30\\[-1.0pt]
52187.2673  & -- 882   & 16 &1072 & 30\\[-1.0pt]
52188.2395  & -- 908   & 18 & 838 & 30\\
\hline
\end{tabular}
\end{table}  }

\section{ Magnetic period of $\gamma$ Equ }
\label{sec:magnet}

Magnetic observations presented in Table~\ref{tab:saores} represent 
completely new data. They cover time span of ca. 7 years 
and include the
phase when the effective magnetic field $B_e$ in $\gamma$ Equ apparently
reached its minimum value, and then the slow decrease of $B_e$ observed
in the recent $\approx$ 50 years has been reversed. This fact is of 
extraordinary importance, because it allows one for a fairly accurate 
determination of the magnetic period and the amplitude of $B_e$ variations
in $\gamma$ Equ.

\begin{figure}
%\resizebox{\hsize}{0.8\hsize}{\rotatebox{0}{\includegraphics{gam_long.eps}}}
\resizebox{\hsize}{0.8\hsize}{\rotatebox{0}{\includegraphics{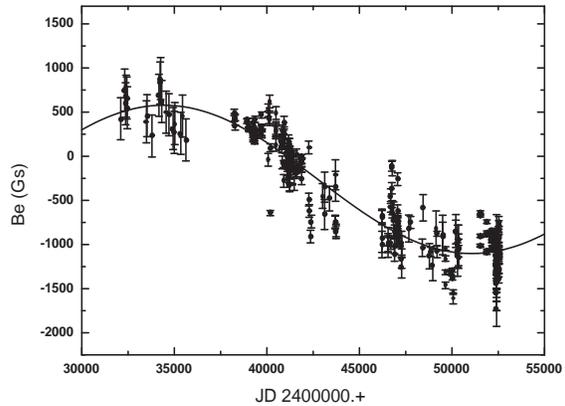}}}
\caption[]{The longitudinal magnetic field $B_e$ for $\gamma$ Equ in years
   1946--2004. }
\label{fig:long}
\end{figure}

We have compiled the set of 298 observations of the $B_e$ field in
$\gamma$ Equ, scattered in the literature, and appended our measurements.
These data cover the time period 1946--2004 (58 years). They are
displayed in Fig.~\ref{fig:long}. 
Note, that the $B_e$ measurements obtained by Babcock (1958) apparently
cover the phase of the maximum longitudinal magnetic field in $\gamma$ Equ.

The set of $B_e$ measurements analysed in this paper is rather 
heterogeneous. The data have been obtained by several different observers
over a long time period using various instruments and techniques, and it
is impossible to estimate or test credibly their systematic and random errors,
particularly for the earliest observations of the longitudinal magnetic
field in $\gamma$ Equ.

Therefore, we arbitrarily assumed that systematic errors of the $B_e$
observations are equal to zero. In other words, all the $B_e$ points for
$\gamma$ which were found in the literature are fully compatible.

Random errors of individual $B_e$ points frequently were given in the
source papers, and are denoted by vertical bars in Fig.~\ref{fig:long}.
These errors
were not directly available for the earliest 
photographic measurements by H.W. Babcock (1958) and Bonsack \& Pilachowski 
(1974). We adopted here the estimated error for Babcock's data equal
238 G, and 151 G for Bonsack \& Pilachowski. These numbers were obtained 
in our thorough reanalysis of the earliest papers dealing with measurements
of stellar magnetic fields, cf. Section 3.1 in Bychkov et al. (2003).

Determination of the period and other parameters of the apparent magnetic
variability for $\gamma$ Equ was performed in the following manner.
Assuming that the run of the observed longitudinal field $B_e$ with time
$T$ can be approximated by a sine wave 
\begin{equation}
  B_e (T)=B_0+B_1 \sin\left[{2\pi (T-T_0)\over P}-{\pi\over 2}\right] \, ,
\label{equ:sigma1}
\end{equation}    
we determined all four parameters: the period $P$, the average field $B_0$,
the amplitude $B_1$ and the time of zero phase $T_0$
using the iterative technique of nonlinear fitting.

Starting values of $P$, $B_0$, $B_1$, $T_0$ and their standard deviations
were found by our computer code for the nonlinear least squares method
(Bychkov et al. 2003). 
The final values and their errors were then computed with the public domain
code ``nlfit.f'', which is designed for curve and surface fitting with the 
Levenberg-Marquardt procedure ({\sc ODRPACK v. 2.01} subroutines). The code
is available at the site {\tt www.netlib.org}.

% 1.01.1946 (12 h UT) = JD 2431822.0
% 1.01.1947 (12 h UT) = JD 2432187.0

Fitting of a sine wave to all the 298 $B_e$ points with errors as in
Fig.~\ref{fig:long} gave very poor results with the $\chi^2$ for
a single degree of freedom $\chi^2/\nu = 18.0420$. Such fits are unacceptable,
and in case of $\gamma$ Equ the poor fit is the result of underestimated
errors of many $B_e$ points. Many $B_e$ observations presented in 
Fig.~\ref{fig:long} have very low errors, which sometimes are less than 20 G.
Our new $B_e$ points, which are collected in Table~\ref{tab:saores}, also
are of such a high formal accuracy.

We cannot judge, whether an apparent scatter of $B_e$ points in
Fig.~\ref{tab:saores} is due to unrealistic error estimates or the intrinsic
short-term variability of the longitudinal magnetic field in $\gamma$ Equ.
The estimated random error of $B_e$ points about the starting sine wave
equals to 213 G. For the final fitting of a sine we assumed that all the
298 points have identical errors of 213 G. 

Final
values of the fitted parameters and their standard deviations $\sigma$
for the sine phase curve are given below. 
\halign {\hskip 1 cm #\hfil\hskip1mm &#\hfil \cr
\noalign{\vskip 5 mm}
   $P_{mag}$  & = $33278 \pm 1327$ days $= 91.1 \pm 3.6$ years \cr
   $T_0 $     & = JD $2417795.0 \pm 1057. $      \cr
   $B_0 $     & = $-\, 262 \pm 22.4 $ G          \cr
   $B_1 $     & = $+\, 839  \pm 22.1 $ G         \cr
   $r   $     & = $-\, 0.524 \pm 0.043$          \cr
\noalign{\vskip 5 mm}
}
\noindent
In other words, a parameter range from $-\sigma $ to $+\sigma$ is just
the true 68\% confidence interval for this parameter. 

The above fit of a sine wave with uniform errors of 213 G is very good, with
$\chi^2/\nu = 1.0134$. The effect of inhomogeneity in the $B_e$ time series
plus the possible existence of rapid magnetic variability in $\gamma$ Equ
were compensated by the increase of the random error, and neither should
influence the above parameters of secular magnetic variability in 
$\gamma$ Equ.

The standard parameter $r$ was defined for the oblique rotator model
of an Ap star. It is related to the angle $\beta$ between the magnetic
dipole axis and the rotational axis, and the angle $i$ between the rotational
axis and the line of sight (Preston 1967):
\begin{equation}
r = {{\cos\beta \cos i - \sin\beta \sin i} \over
     {\cos\beta \cos i + \sin\beta \sin i}} 
  = {B_e (\min) \over {B_e (\max)}}  \, .
\label{eqn:rrr}
\end{equation}
Parameters $B_e (\min)$ and $B_e (\max)$ of the $B_e$ sine wave for 
$\gamma$ Equ are given by
\halign {\hskip 1 cm #\hfil\hskip1mm &#\hfil\hskip1mm &# \hfil \cr
\noalign{\vskip 2 mm}
   $B_e({\rm max}) $ & $=B_0+B_1$ & = $+\,\,\, 577 \pm 31.4$ G   \cr
   $B_e({\rm min}) $ & $=B_0-B_1$ & = $ -1101 \pm 31.4$ G        \cr
\noalign{\vskip 2 mm}
}
Note, that the meaning of $B_e({\rm max}) $ and $B_e({\rm min}) $ for
use in Eq.~\ref{eqn:rrr} is different: $B_e({\rm max}) $ denotes there
the value of magnetic intensity which has the higher absolute value, and
$B_e({\rm min})$ has the lower absolute value. In this way we obtained the
value of $r$ for $\gamma$ Equ equal to $r=577 / (-1101) = -0.524 $.

Bychkov et al. (2005a) presented an extensive catalog of the magnetic
phase curves and their parameters for 136 stars on the main sequence and
above it. We quoted there the 
previously estimated period for $\gamma$ Equ, $P_{mag}=27027^d$,
which was obtained on the basis of a shorter series of $B_e$ data.
This paper and the new, more accurate $P_{mag} = 33278^d$ represents a 
major revision of the previously known magnetic period of $\gamma$ Equ.

\section{Search for additional magnetic periods in $\gamma$ Equ}

Significant scatter of the observed points in the long-term run of $B_e (T)$
in Fig.~\ref{fig:long} suggests the search for short-term periodicities.
We applied the strategy of prewhitening to the set of available $B_e$
measurements, and removed the principal sine-wave variations from the data.
Prewhitened data were then analysed with the method developed by Kurtz 
(1985), and with his Fortran code (Kurtz 2004). 

Such a search for peaks in the $B_e$ amplitude spectrum of $\gamma$ Equ
in this paper was restricted to trial periods higher than 1 day. 
This is because many of the earlier magnetic observations for this star 
either have poorly determined the time of measurement, or have 
long times of exposure (see e.g. Babcock 1958). 
The star $\gamma$ Equ exhibits rapid nonradial pulsations and the
corresponding $B_e$ with the period $P_{mag}=12.1$ min (Leone \& Kurtz
2003) and, possibly, with simultaneous shorter periods 
(Bychkov et al. 2005b). None of them were analysed in this paper.

\begin{figure}
\resizebox{\hsize}{0.8\hsize}{\rotatebox{0}{\includegraphics{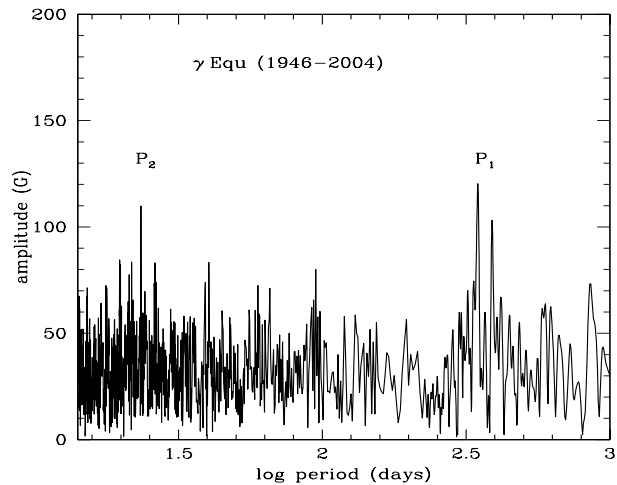}}}
\caption[]{Amplitude spectrum of the $B_e$ time series for 
   $\gamma$ Equ, years 1946--2004. }
\label{fig:short}
\end{figure}

We have identified two additional periods of statistically low significance
in the range $P_{mag} > 1^d$, see Fig.~\ref{fig:short}:

\vskip 3 mm
$P_1 = 348.07$ days, amplitude $=122$ G   \par
$P_2 =  23.44$ days, amplitude $=110$ G   \par
\vskip 3 mm

\noindent
Both peaks in the amplitude spectrum in Fig.~\ref{fig:short} exhibit low
signal to noise ratio, with noise level at ca. 80 G. The period $P_1$ 
is close to 1 year. Since most of the existing $B_e$ observations for $\gamma$
Equ were performed in months July-November, then the peak $P_1$ in the
amplitude spectrum represents a false period which most likely reflects the
average 1-year repetition time in the acquisition of the existing magnetic
measurements.

We believe that the peak $P_2$ in the amplitude spectrum of the $B_e$ field
of $\gamma$ Equ is the random effect of a pure noise. The peak is very
narrow, in fact, it only appears in a single bin of a very dense discrete
frequency mesh.

Kurtz (1983) discussed the possible existence of the period of $\approx 38$
days in his photometric observations of $\gamma$ Equ
in 1981. That period was of low probability, but  
possibly could be identified with the real rotational period in
this star. We do not confirm the existence of the 38 day period in 
long-term $B_e$ observations of $\gamma$ Equ, see Fig.~\ref{fig:short}.

% 1 day^-1  = frequency 0.01157408 mHz
% 1 year^-1 = frequency 3.170e-05 mHz

\section{ Discussion }
\label{sec:discussion}

There exist three possible explanations for the observed long-term behavior 
of the longitudinal magnetic field in $\gamma$ Equ:

\vskip 3 mm
\noindent
{\bf 1.} Precession of the rotational axis (Lehmann 1987).   \par
\noindent
{\bf 2.} Solar-like magnetic cycle (Krause \& Scholz 1981),  \par
\noindent
{\bf 3.} Rotation with the period of 91.2 years.   
\vskip 3 mm

The Ap star $\gamma =$ HD 201601 in fact is a binary system. One can
assume, that the gravitational force from the secondary companion can cause
precession of the Ap star. As the result, the angle between the rotational
axis and the direction towards the Earth varies periodically. Therefore,
changes of the aspect can in principle cause apparent variations of the
longitudinal magnetic field $B_e$ or the amplitude of its variations.

Effects of precession in long-period Ap stars were studied by Lehmann 
(1987), who showed that the oblateness of stars caused by the rotational
or magnetic flattening is not adequate to produce observable precession 
effects. The only exception was 52 Her, where the observed behavior of 
the star could be interpreted as a precessional motion. 

The above considerations indicate that the precession theory does not
convincingly explain $B_e$ variations in this star. 

The idea by Krause \& Scholz (1981) that we actually observe the solar-like
magnetic cycle in $\gamma$ Equ in which the global magnetic field
reverses its polarity, cannot be easily verified by the existing 
observations of the global longitudinal magnetic field $B_e$. Moreover,
one can note that such an idea requires the existence of a mechanism in the
interior of $\gamma$ Equ which ensures the transfer of huge magnetic
energy into electric currents and vice versa. Note that the required
efficiency of such a mechanism and the amplitude of magnetic field 
variations in $\gamma$ Equ is ca. four orders of magnitude larger than
that in the Sun in a similar timescale.

Following the widely accepted picture of an Ap
star, we believe that the magnetic field of $\gamma$
Equ can be approximated by a dipole located in the center
of the star. The dipole is inclined to the rotational axis of $\gamma$ Equ.
We assume that the magnetic field is stable and remains frozen in the 
interior of a rotating star at the time of observations, i.e. during at
least of 58 years. Therefore, slow variations of the $B_e$ field
in $\gamma$ Equ are caused by an extremely slow rotation, in which case our
$P_{mag} = P_{rot} = 33278^d$. Such an explanation is supported to some 
extent by polarimetric measurements by Leroy et al. (1994).

We plan to perform high accuracy polarimetric 
measurements of $\gamma$ Equ with the new version of MINIPOL. The device
was constructed to measure the angles and the degree of linear polarisation
of stellar radiation, and will be operational at the Special
Astrophysical Observatory in 2006. We also expect that shall be able to
verify the extremely slow rotation of $\gamma$ Equ measuring the rate of 
change for the polarisation angle of stellar radiation.

\section{Summary }

The Ap star $\gamma$ Equ (HD 201601) exhibited slow and systematic
decrease of the longitudinal magnetic field $B_e$ starting from 1946,
when the global magnetic field of this star was discovered (Babcock 1958).
We have compiled the full set of 298 existing $B_e$ measurements, which
consists of the $B_e$ data published in the literature and our observations
obtained during recent 7 years. The latter magnetic data (33 $B_e$ points)
were measured with the echelle spectrograph in the Coude focus of the 1-m
telescope at the Special Astrophysical Observatory.
Our newest observations showed that the longitudinal magnetic field $B_e$
of $\gamma$ Equ reached its local minimum and started to rise in 1998-2004.

All the available data cover the time period of 58 years (1946-2004) and
include both phases of the maximum and minimum $B_e$. 
Assuming that the secular variability of the $B_e$ field is a periodic
feature, we determined parameters of the magnetic field curve in
$\gamma$ Equ and give the value of its period, $P=91.1 \pm 3.6$ years,
with the zero phase (maximum of $B_e$) at 
$T_0 =$ JD $2417795.0 \pm 1057$. Sine-wave fit to the $B_e$ phase curve 
yields $B_e({\rm max}) =+577 \pm 31$ G and $B_e({\rm min}) =-1101 \pm 31$ G.

Spectral analysis of the 58-year long $B_e$ time series essentially do not
show the existence of shorter periods, down to trial periods of $\approx$
1 day. More specifically, there are no real shorter periods in the run of the
longitudinal magnetic field $B_e$ with amplitudes exceeding the noise
level of 80 G.

\section*{Acknowledgments}

We are grateful to John D. Landstreet, the referee, for his criticism
and suggestions regarding our computations and the manuscript. 
We thank Don Kurtz for providing his Fortran software used
here to compute the amplitude spectrum of $\gamma$ Equ.  
This research was supported by the Polish Committee for Scientific
Research grant No. 1 P03D 001 26.

% for providing us the fortran code nlfit.f we used in our reseach.


\begin{thebibliography}{}

\bibitem[\protect\citeauthoryear{}{}]{} 
Babcock, H.W., 1958, ApJS, 3, 141

\bibitem[\protect\citeauthoryear{}{}]{}                                                        
Bonsack, W.K., Pilachowski, C.A., 1974, ApJ, 190, 327

\bibitem[\protect\citeauthoryear{}{}]{}
Borra, E.F., Landstreet, J.D., 1980, ApJS, 42, 421  

\bibitem[\protect\citeauthoryear{}{}]{}
Bychkov, V.D., Shtol', V.G., 1997, Stellar Magnetic Fields,
   Proc. Int. Conf., 200

\bibitem[\protect\citeauthoryear{}{}]{}
Bychkov, V.D., Fabrika, S.N., Shtol', V.G., 1991, Pis'ma Astron. Zh., 17, 43

\bibitem[\protect\citeauthoryear{}{}]{}
Bychkov, V.D., Romanenko, V.P., Bychkova, L.V. 1998, Bull. Spec. Astrophys.
   Obs., 45, 110

\bibitem[\protect\citeauthoryear{}{}]{}
Bychkov, V.D., Romanenko, V.P., Bychkova, L.V. 2000, Bull. Spec. Astrophys.
   Obs., 49, 147

\bibitem[\protect\citeauthoryear{}{}]{}
Bychkov, V.D., Bychkova, L.V., Madej, J., 2003, A\&A 407, 631

\bibitem[\protect\citeauthoryear{}{}]{}
Bychkov, V.D., Bychkova, L.V., Madej, J., 2005a, A\&A 430, 1143

\bibitem[\protect\citeauthoryear{}{}]{}
Bychkov, V.D., Bychkova, L.V., Madej, J., 2005b, Acta Astron., 55, 141

\bibitem[\protect\citeauthoryear{}{}]{}
Hildebrandt, G., Scholz, G., Lehmann, H., 2000, Astron. Nachr., 321, 115

\bibitem[\protect\citeauthoryear{}{}]{}
Hubrig, S., Kurtz, D.W., Bagnulo, S., Szeifert, T., Sch\"oller, M., 
   Mathys, G., Dziembowski, W.A., 2004, A\&A, 415, 661

\bibitem[\protect\citeauthoryear{}{}]{}
Krause, F., Scholz, G., 1981, Chemically peculiar stars of the upper main
   sequence, 23rd Int. Conference on Astrophys., Liege, Belgium, 323

\bibitem[\protect\citeauthoryear{}{}]{}
Kurtz, D.W., 1983, MNRAS, 202, 1

\bibitem[\protect\citeauthoryear{}{}]{}
Kurtz, D.W., 1985, MNRAS, 213, 773

\bibitem[\protect\citeauthoryear{}{}]{}
Kurtz, D.W., 2004, personal communication

\bibitem[\protect\citeauthoryear{}{}]{}
Lehmann, H., 1987, Astron. Nachr., 308, 333

\bibitem[\protect\citeauthoryear{}{}]{}
Leone, F., Kurtz, D.W., 2003, A\&A, 407, L67

\bibitem[\protect\citeauthoryear{}{}]{}
Leroy, J.L., Bagnulo, S., Landolfi, M., Degli'Innocenti, E. Landi,
   1994, A\&A, 284, 174

\bibitem[\protect\citeauthoryear{}{}]{}              
Mathys, G., 1991, A\&AS, 89, 121

\bibitem[\protect\citeauthoryear{}{}]{}
Mathys, G., Hubrig, S., 1997, A\&AS, 124, 475

\bibitem[\protect\citeauthoryear{}{}]{}
Monin, D.N., 1999, Bull. Spec. Astrophys. Obs., 48, 121

\bibitem[\protect\citeauthoryear{}{}]{}
Musaev, F.A., 1996, Astronomy Letters, 22, 715

\bibitem[\protect\citeauthoryear{}{}]{}
Preston, G.W., 1967, ApJ, 150, 547

\bibitem[\protect\citeauthoryear{}{}]{}
Scholz, G., 1975, Astron. Nachr., 296, 31

\bibitem[\protect\citeauthoryear{}{}]{}
Scholz, G., 1979, Astron. Nachr., 300, 213

\bibitem[\protect\citeauthoryear{}{}]{}
Scholz, G., Hildebrandt, G., Lehmann, H., Glagolevskij, Yu.V.,
   1997, A\&A, 325, 529

\bibitem[\protect\citeauthoryear{}{}]{}
Zverko, J., Bychkov, V.D., Ziznovsky, J., Hric, L., 1989, Contr. Astron.
    Observatory Skalnate Pleso, 18, 71

\end{thebibliography}
\end{document}